\documentclass[iop, apj]{emulateapj}
\usepackage[colorlinks,urlcolor=blue,citecolor=blue,linkcolor=blue]{hyperref}
\usepackage{color}

\newcommand{\herschel}{\textit{Herschel}}
\newcommand{\swift}{\textit{Swift}}

\begin{document}

\title{Do Most Active Galactic Nuclei Live in High Star Formation Nuclear Cusps?\footnotemark[1]}
\author{Richard F. Mushotzky\altaffilmark{2}, T. Taro Shimizu\altaffilmark{2}, Marcio Mel\'endez\altaffilmark{2}, and Michael Koss\altaffilmark{3}}
\footnotetext[1]{{\it Herschel} is an ESA space observatory with science instruments provided by European-led Principal Investigator consortia and with important participation from NASA.}
\altaffiltext{2}{Department of Astronomy, University of Maryland, College Park, MD 20742, USA}
\altaffiltext{3}{Institute for Astronomy, University of Hawaii, Honolulu, HI 96822, USA}
\email{richard@astro.umd.edu}

\begin{abstract}
We present early results of the Herschel PACS (70 and 160 \micron{}) and SPIRE (250, 350, and 500 \micron{}) survey of 313 low redshift ($\rm{z} < 0.05$), ultra-hard X-ray (14--195 keV) selected AGN from the 58 month Swift/BAT catalog. Selection of AGN from ultra-hard X-rays avoids bias from obscuration providing a complete sample of AGN to study the connection between nuclear activity and star formation in host galaxies. With the high angular resolution of PACS, we find that $>$35\% and $>$20\% of the sources are ``point-like'' at 70 and 160 \micron{} respectively and many more that have their flux dominated by a point source located at the nucleus. The inferred star formation rates (SFR) of 0.1 - 100 M$_{\sun}$ yr$^{-1}$ using the 70 and 160 \micron{} flux densities as SFR indicators are consistent with those inferred from Spitzer NeII fluxes, but we find that 11.25 \micron{} PAH data give $\sim$3x lower SFR. Using GALFIT to measure the size of the FIR emitting regions, we determined the SFR surface density [M$_{\sun}$ yr$^{-1}$ kpc$^{-2}$] for our sample, finding a significant fraction of these sources exceed the threshold for star formation driven winds (0.1 M$_{\sun}$ yr$^{-1}$ kpc$^{-2}$).
\end{abstract}


\section{Introduction}\label{intro}
Over the past 20 years there have been many indications of a strong connection between star formation and nuclear activity . \citep[e.g.][]{Alexander:2005ly, Schweitzer:2006mz, Netzer:2007ve, Netzer:2009lr, Diamond-Stanic:2012rw, Rosario:2012fr, Santini:2012ly}. However much of this work is based on observations of moderate to high redshift optically or IR selected systems in which it is difficult, because of a lack of angular resolution, to connect the infrared (IR) radiation to star formation or active galactic nuclei (AGN) processes. Optically selected samples are also strongly biased with respect to dust, AGN absorption and other galaxy host properties \citep{Mushotzky:2004gf}. 

Because the nuclear light from the AGN frequently makes analysis of the stellar population difficult, many authors have focused on the analysis of type II AGN and thus have had to assume that the connection between the AGN and the host galaxy are similar in type Is and IIs. It is not clear that such an assumption is correct since recent results \citep{Donoso:2013ve} show that obscured sources have a higher clustering signal than unobscured sources and thus lie in more massive objects than unosbscured sources. Thus the nature of the host galaxies of type II AGN are likely to be rather different than that of type Is. 

Since, as we will show in this paper, the IR radiation in low-z AGN is strongly nuclear in nature, the assumption that commonly used star formation  templates are sufficiently accurate to model the star formation in AGN ignores the possibility that the AGN may influence the spectral nature of the star formation \citep[e.g.][]{Meijerink:2013qy}. Many authors have made the strong assumption that the IR continuum, originates solely from from torus \citep{Nenkova:2002ys, Fritz:2006yq, Schartmann:2008vn, Lira:2013rt} to model the IR emission from the AGN. Such assumptions can strongly effect the interpretation of the far-IR (FIR) radiation. 

To understand the relation of star formation to AGN activity, the bolometric luminosity of the AGN, and the nature of the IR continuum we have conducted a \herschel{} \citep{Pilbratt:2010rz} program of broad band (70-500\micron{}) imaging observations of 320 ultra hard (14-195 keV) X-ray selected AGN from the 58 month Swift BAT catalog\footnote{\url{http://heasarc.nasa.gov/docs/swift/results/bs58mon/}} \citep{Tueller:2010ul}. Because of the essentially flux limited nature of this low redshift survey the luminosities of the objects are strongly correlated with redshift and span an ultra hard X-ray luminosity range from 41.5-45.4 ergs/sec. The vast majority of these objects are Seyfert galaxies, with very few LINERS or other extremely low luminosity AGN or very high luminosity quasars. 

The \swift{} Burst Alert Telescope (BAT) \citep{Barthelmy:2005ul} sample is selected entirely as a signal to noise limited catalog of sources detected in the 14-195 keV and identified with active galaxies via follow-up or archival observations. This sample is not biased with respect to the optical, IR or radio properties of the host and has a very high identification fraction \citep{Baumgartner:2012gf}. However, it is biased against so-called Compton thick AGN in which the line of sight column density is greater than $2\times10^{24}\,\,\rm{cm}^{-2}$. We have applied a low redshift cutoff of $\rm{z} < 0.05$ in order to obtain interesting constraints on the spatial origin of the FIR emission for the first time given \herschel's angular resolution of ~5.8" full-width half-max (FWHM) for its 70 \micron{} waveband. Details of the sample selection and observations will be discussed in Mel\'endez et al 2014, in prep. We obtained a completeness fraction of 94\% at 70 \micron{}, 80\% at 160 \micron{}, 85\% at 250 \micron{}, 68\% at 350 \micron{} and 46\% at 500 \micron{} with 5$\sigma$ confidence giving 296 sources at 70 \micron{} and 259 at 160 \micron{}. 

In this paper we focus on the higher angular resolution PACS \citep{Poglitsch:2010fp} data at 70 \micron{} and 160 \micron{} and leave to further work detailed modeling of the broad band spectra and correlations with other wavebands (Melend\'ez et al 2014 in prep). 

\section{Observational Details}\label{obs_details}
The vast majority of the BAT AGN presented in this work are from our Cycle 1 open-time program (OT1\_rmushotz\_1, PI: R. Mushotzky) with a total of 291 sources. We included an additional 22 BAT sources from different programs publicly available from the Herschel science archive (HSA), see Mel\'endez et al 2014  for details, giving a total number of 313 sources in our sample. For the sources obtained through our OT1 program the PACS imaging for the blue 70 \micron{} (60-85 \micron) and red 160 \micron{} (130-210 \micron) band  was obtained simultaneously  in mini-scan mode along two  scan map position angles at  70 and 110 degrees. Each orientation angle with a  medium scan speed of 20" s$^{-1}$,   2 scan legs of 3.0' length with 4.0" scan leg separation and a repetition factor of 1. The total time per observation was 52 s\footnote{From our OT1 program II SZ 010, Mrk 290, PG 2304+042 and Mrk 841 have a different configuration with 10 scan legs of 3.0' length with 4.0" scan leg separation and a repetition factor of 1 with a total time per observation of 276 s}. 

For the PACS data reduction we used the Herschel Interactive Processing Environment \citep[HIPE,][]{Ott:2010rm} version 8.0. The "Level 0" observations (raw data)  were processed through the standard pipeline procedures. To correct for the bolometers drift (low frequency noise), both thermal and non-thermal (uncorrelated noise), and to create the final maps  we used the algorithm implemented in Scanamoprhos \citep[v19.0,][]{Roussel:2012vn} .

PACS fluxes are measured  through aperture photometry. Circular apertures of varying radii were placed to measure the source flux and concentric annuli were used to estimate the background flux. For the more extended  sources, we used elliptical apertures and annuli. The apertures  are chosen by eye to contain all of the observed emission at each wavelength and the background annulus was set to encompass a clean, uncontaminated sky region close to the source. Finally,  aperture corrections were applied to account for flux outside the source apertures.

\section{Results}
\subsection{Point Source Contribution}
Visually, it is easy to see that for a large fraction of our sample, the majority of the 70 \micron{} emission originates from a small point-like component. To estimate the contribution from a centrally located point source, we extracted fluxes from a small 6" aperture for the PACS 70 \micron{} images, and a 12" aperture for the 160 um image. The sizes of the apertures are roughly the FWHM's of the PSFs in each waveband to minimize contribution from any extended component. These point source fluxes are then compared to the global FIR flux as determined by Mel\'endez et al 2014 (in prep).

\begin{figure}
\plotone{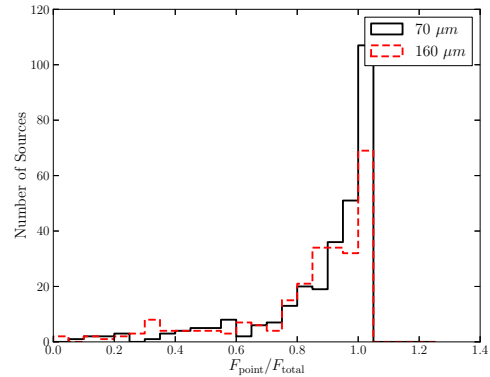}
\caption{Distribution of the point source contribution to the 70 and 160 \micron{} fluxes. Point source fluxes were extracted from a 6" and 12" aperture for 70 and 160 \micron{} respectively. Nearly all of the sources in the sample have a point source contribution $>50\%$ .\label{pnt_src_cont}}
\end{figure}

Figure \ref{pnt_src_cont} shows the distribution of point source contributions for both 70 and 160 \micron{}. We find that 274/296 (92.5\%) of the objects have a point source contribution greater than 50\% of the total flux at 70 \micron{} and 229/259 (88.4\%) at 160 \micron{}. The remaining sources are mainly large, extended very nearby galaxies.

\subsection{FIR Sizes}
Using GALFIT \citep{Peng:2002dq}, we estimated the FIR size of our sources by modeling the FIR surface brightness with a 2D Gaussian. Because the shape of the PACS point spread function (PSF) is dependent on the details of the observation (i.e. scan angles, scan speed, etc.) and mapmaking technique, we downloaded and reduced an observation of one of the calibration stars, $\alpha$ Tau (OBSID 1342240755 and 1342240756). $\alpha$ Tau was observed with the same scan speed and scan angles as our sample, and we used the same scripts and Scanamorphos version to reduce the data and create a map. $\alpha$ Tau was then cut out of the map and input into GALFIT as the instrument PSF to be convolved with the chosen model. Through visual inspection of the residual images, it seems that,  for 105/296 sources at 70 \micron{} (50/259 at 160 \micron{}) only a point source component is necessary to fit the surface brightness. This was verified by a detailed analysis which  determined an upper limit on the size for these sources. This size was obtained by iteratively increasing the FWHM parameter for the Gaussian model until we achieved a 3$\sigma$ change in $\chi^{2}$, determining the largest a source could physically be but still be unresolved for PACS. 

For another 106 objects at 70 \micron{} (128 at 160 \micron), the sources are quasi-resolved and a 2D Gaussian model fit the images well. We utilize the best-fit FWHM as the radius for the source size. We also tested an exponential profile to these sources at 70 \micron{} since this has been found to fit star-forming disks. Both the Gaussian and exponential profile are special cases of the more general S\'ersic profile with each having specific values of the ``compactness" parameter, $n$ ($n=0.5$ and $n=1$ respectively). However all three profiles have different ``size" parameters that are related to each other: effective radius, $r_e$ for the S\'ersic, scale length, $r_s$ for the exponential, and FWHM for the Gaussian. To fairly compare the sizes found by the exponential and Gaussian profiles, we scaled each to $r_e$ using the following relations.

\begin{equation}
r_e = 1.678r_s
\end{equation}
\begin{equation}
r_e  = 0.588FWHM
\end{equation}
The median ratio of $r_e$ determined from the exponential to the Gaussian is 0.76 meaning an exponential profile finds even smaller effective sizes for our sources and using a Gaussian is likely a conservative model to measure the FIR size of our sources. For the rest of the paper we use the sizes found with the Gaussian model.

The rest of the sources in the sample are clearly extended and either have non-Gaussian morphologies (e.g. spiral or clumpy structure) and are too well resolved to be fit by this simple model. For this subsample, we use the aperture radius implemented to measure the global flux as a reasonable estimate of the size of the source. These should be at worst conservative upper limits since apertures were chosen by eye to fully encompass the FIR emission of each source.

\begin{figure}
\begin{center}
\plottwo{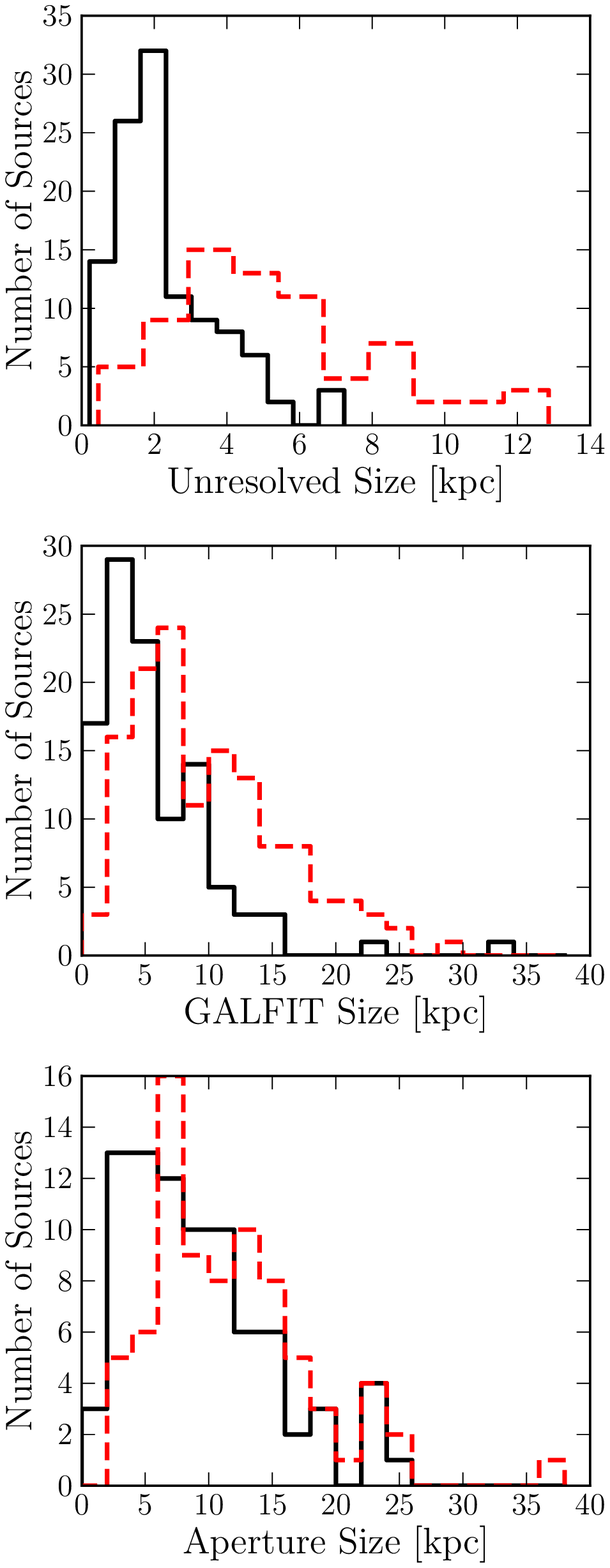}{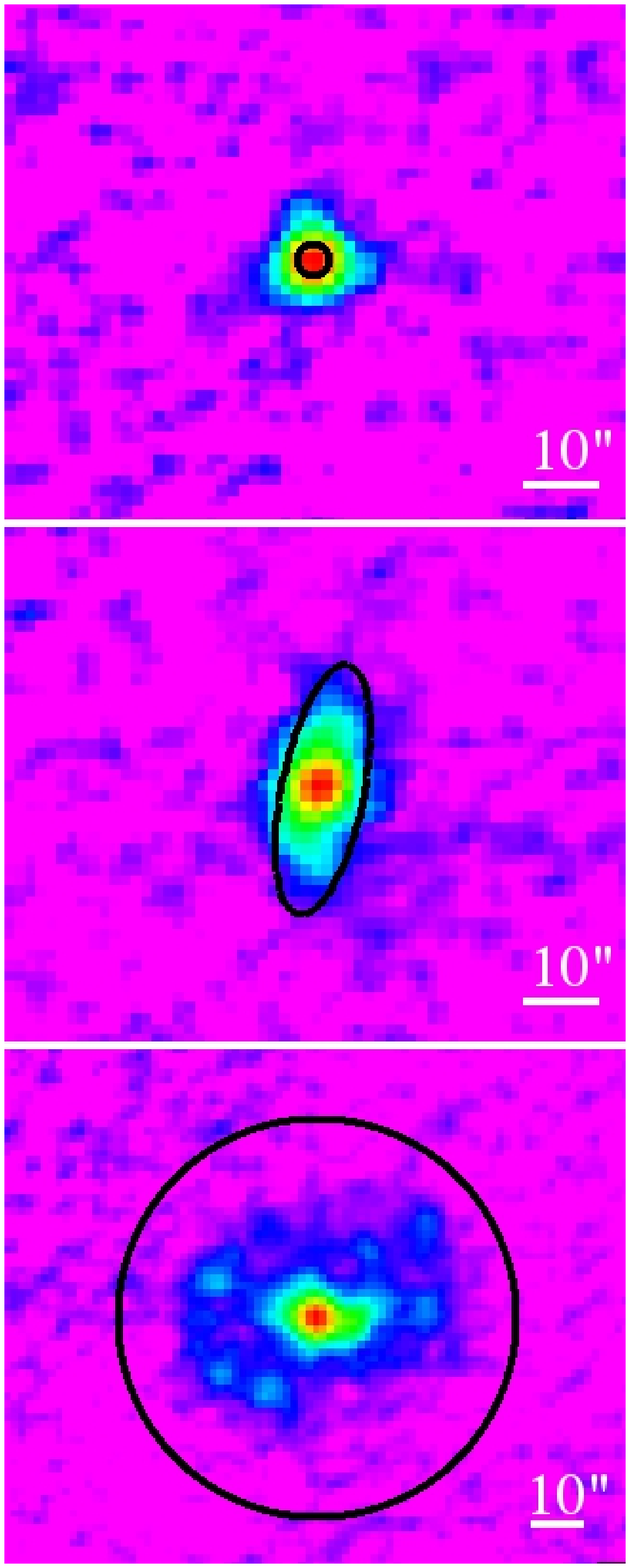}
\caption{Distribution of measured sizes and example sources for the point source sample (top), sample measured with GALFIT (middle), and sample with sizes equal to the aperture size (bottom). Each histogram contains two distributions, one for 70 \micron{} (solid black) and one for 160 \micron{} (dashed red). Each image is displayed with asinh scaling from a flux level of 0 to the maximum flux of the source. The black circles/ellipses outline the measured size used for the source. From top to bottom the specific AGN are Mrk 704, ESO 197-G04, and NGC 5674.\label{src_sizes}}
\end{center}
\end{figure}

In Figure \ref{src_sizes} we show the distribution of source sizes with upper limits representing the 'point-like' objects. We find a range of upper limits on the source size from a median value of 3.1" at 70 \micron{} and 7.7" at 160 \micron{} or a median value of 2 kpc at 70 \micron{} and 5 kpc at 160 \micron{}. Thus for the vast majority of the hard X-ray selected AGN the dominant far-IR component is nuclear in nature which is the same as what has been found in (U)LIRGS in the mid-IR \citep{Diaz-Santos:2010vn}).

One possible explanation for the compactness of our sources is low sensitivity to faint extended emission due to the short exposures. We tested this by comparing the radial profiles of 6 sources that have duplicate longer exposure (at least a factor of 4 longer) observations in the {\it Herschel} archive. These observations were reduced in the same manner as our sample and the same mapmaking routine was used to create the maps. Comparison of the radial profiles shows no significant extended emission being missed in our short exposures in 5/6 sources out to radii of 80", the shortest length of our images. The one source with ``extra" extended emission in the longer exposure image, NGC 7465, seems to show a tidal tail due to interaction with a nearby companion, and is not related to star formation within the galaxy. We conclude that our short observations are not missing significant faint extended emission.

\subsection{Star Formation Rates}
\begin{figure}
\begin{center}
\epsscale{0.5}
\plotone{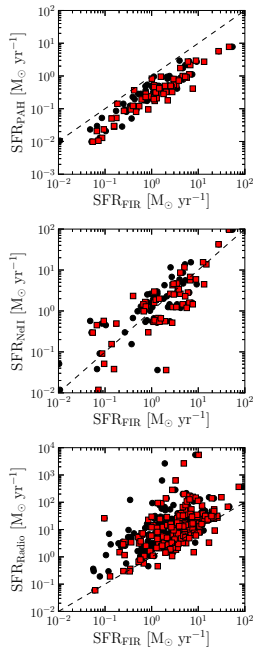}
\end{center}
\caption{Comparison of the 5 different SFR indicators (70 \micron{}, 160 \micron{}, NeII, PAH 11.25 \micron{}, 1.4 GHz). The black circles are 70 \micron{} determined SFR while the red squares are 160 \micron{} determined. The black dashed lines are a 1:1 correspondence between PAH 11.25 \micron{}(top)/NeII (middle)/1.4 GHz (bottom) with a FIR determined SFR.\label{sfr}}
\end{figure}

To estimate the star formation rate (SFR) of our sources, we first used the monochromatic calibration of 70 and 160 \micron{} from \citet{Calzetti:2010qf}. We compared these SFRs to those measured by the near IR NeII (108 galaxies) and PAH 11.25 \micron{} (79 galaxies) emission extracted from Spitzer IRS spectra as well as 1.4 GHz radio fluxes (189 galaxies) from the NVSS catalog. For the [Ne~II] flux we performed a line fit with a polynomial to fit the continuum and a Gaussian for the line profile \citep{Weaver:2010rt}. The 11.25 \micron{} PAH feature was measured by integrating the flux above a spline-interpolated continuum (see \citet{Spoon:2007zr} for details on the procedure). [Ne~II] SFR's were calculated from the calibration in \citet{Melendez:2008pd} corrected for a \citet{Kroupa:2001qy} initial mass function (IMF) while for the PAH 11.25 \micron{} SFR we used the calibration from \citet{Diamond-Stanic:2012rw}. For 1.4 GHz we used the \citet{Murphy:2011rt} calibration. We find (Figure \ref{sfr}) that the far-IR inferred star formation rates of 0.1 - 100 M$_{\sun}$yr$^{-1}$ are consistent with those inferred from Spitzer NeII fluxes but with wide scatter, while the PAH 11.25 \micron{} derived SFR are systematically ~3x lower and the 1.4 GHz relation gives systematically higher SFRs by a factor of 2-4. The discrepancy between the radio derived and FIR derived SFRs might be explained due to the presence of AGN related emission in the radio, but the discrepancy in the PAH derived SFRs seem to suggest that PAHs are being destroyed by the AGN or there is an error in either the NeII or PAH calibrations.

\begin{figure}
\begin{center}
\epsscale{1.0}
\plotone{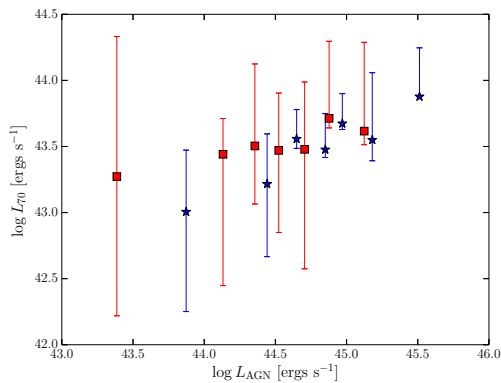}
\end{center}
\caption{70\micron{} luminosity plotted against AGN luminosity which is calculated directly from BAT luminosity. Blue stars and red squares represent Seyfert 1's and 2's respectively. Each star or square is the average $\log L_{70\micron}$ of 20 sources binned in $\log L_{\rm{AGN}}$ with error bars representing the range containing 68\% of the sources in the bin. \label{lum_agn_70}}
\end{figure}

The 70 \micron{} SFRs are roughly a factor of ~2  greater than the 160 \micron{} rates if we use the  \citet{Calzetti:2010qf} calibration derived for starburst galaxies. If we renormalize the 2 rates, such that they both give the same SFR,  the variance between the two estimators  is still a factor of 2. While all the star formation rate indicators are highly correlated, the wide range in inferred rates does not allow us to definitively assign all of the emission to star formation and thus the inferred rates may be significantly (factor of 3) affected by AGN emission. This can also be seen in Figure \ref{lum_agn_70} where a weak correlation between the AGN luminosity, inferred by the BAT luminosity ($L_{\rm{AGN}} \approx 15L_{\rm{BAT}}$ \citep{Winter:2012yq}), and the 70\micron luminosity which is being used as a tracer for star formation, is detected, especially for the Seyfert 1's in our sample.

\subsection{Star Formation Surface Density}
\begin{figure}
\epsscale{1.0}
\begin{center}
\plotone{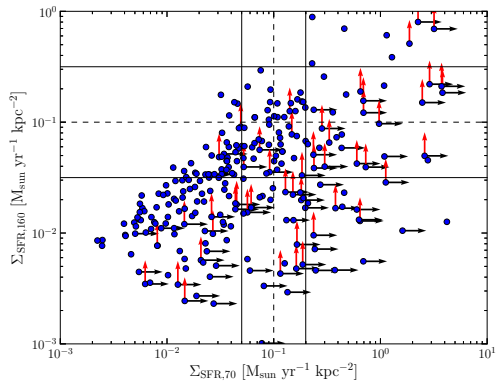}
\end{center}
\caption{Comparison between the SFR surface density from the 70 \micron{} calibration and the 160 \micron{} calibration. Arrows correspond to lower limits on the SFR surface density. Black dashed lines are drawn at 0.1 M$_{\sun}$ yr$^{-1}$ kpc$^{-2}$, the threshold for star formation driven winds from \citet{Heckman:2001ve}. The solid lines represent the uncertainty of 0.3 (70 \micron{}) and  0.5 (160 \micron{}) dex in the star formation calibrations from \citet{Calzetti:2010qf}.\label{sfr_dens}}
\end{figure}

Since we do not have Spitzer Ne II or PAH data for all of our sample we have used the 70 and 160 \micron{} data to infer the star formation rate surface density. Using these SFR's and estimates of the source size we calculate a star formation rate surface density (M$_{\sun}$ kpc$^{-2}$) for both the 70 and 160 \micron{} data (Figure \ref{sfr_dens}).  At least 30\% and as many as 50\% of the objects using 70 um as the SFR indicator or between 20-30\% using 160 um as the indicator have SFR densities larger than the empirical threshold of 0.1 M$_{\sun}$ yr$^{-1}$ kpc$^{-2}$ needed to drive a wind \citep{Heckman:2001ve}.
There is a very similar distribution in star formation rates for objects which are resolved, partially resolved or unresolved suggesting that the Herschel angular resolution is adequate for identifying nearby objects with high specific star formation rates.   The high rate of star formation surface density (M$_{\sun}$ yr$^{-1}$ kpc$^{-2}$) indicates that AGN very often lie in nuclear starbursts which should drive winds. To our knowledge this is the first indication that Seyfert galaxies should have, frequently, star formation driven nuclear winds.

\section{Discussion}
The general agreement, with large scatter, of the inferred star formation rates using 5 different indicators shows that the assumption that the bulk of the 70 and 160 \micron{} luminosities from our sources is consistent with star formation is reasonable. This discovery has only been made possible by the combination of an AGN sample unbiased with respect to host galaxy properties and the Herschel sensitivity and imaging capabilities.  This concentration is seen in both Type Is and IIs which show little or no color differences in the PACS data (Mel\'endez et al 2014 in prep). 

We are thus led to the conclusion that either a significant fraction of the hosts of low redshift AGN have nuclear starbursts, that a significant fraction of the 70 and 160 \micron{} luminosity is not produced by star formation, or that the normalization of the various indicators of star formation is very different in the nuclear starforming regions of AGN.  A similar conclusion was reached by \citet{Diamond-Stanic:2012nx} and \citet{LaMassa:2013qy} using Spitzer observations over a range of redshifts and inferring sizes and was predicted theoretically by \citet{Ballantyne:2008qy}. However this is the first time, to our knowledge that this result is based on direct imaging of the star formation process and on an AGN sample which has direct measures of the AGN  luminosity and is relatively unaffected by selection effects.

The strongly compact morphology of the low redshift BAT sample is a surprise, given their modest inferred star formation rates and indicates that even at moderate AGN luminosities the IR radiation is strongly nuclear in nature and thus might be connected to the AGN phenomenon. In a comparison with the KINGFISH sample \citep{Kennicutt:2011vn} of normal starforming galaxies with {\it Herschel} PACS observations and similar optical absolute magnitudes ($M_{\rm opt}<-19$) and SFRs, the FIR surface areas, inferred from the apertures used for the global fluxes, are a factor of 6 larger than the FIR sizes of our sample. Comparison of the far IR to hard X-ray luminosities of X-ray selected AGN  \citep{Rosario:2012fr, Mullaney:2012gf} show that the ratio L$_{IR}$/L$_{X}$ increases considerably with red shift. If these high-z objects have the same morphology as our lower redshift sample, which will require ALMA to test, the strong relationship between star formation rate surface density and outflow velocity \citep{Newman:2012lq} would imply that virtually all of the high z AGN would have starburst driven winds. 

\section{Conclusions}
The analysis of the 70 and 160 \micron{} images of a sample of 313 nearby ($z<0.05$)  hard X-ray selected AGN shows that in over 90\% of the sources the bulk of the FIR radiation is point-like at the spatial resolution of Herschel (a median value of 2 kpc FWHM). The inferred star formation rates from a variety of indicators (NeII, 70,160 \micron{}, PAH and radio emission) agree within a scatter of ~4 and are consistent with the idea that at least 30\% of the FIR radiation is due to star formation. If the FIR is tracing nuclear star formation, then this is also tracing the cold molecular gas that could be fueling the AGN as suggested by \citet{Hopkins:2013yg}. The combination of the star formation rates and the upper limits on source size shows that at least 30\% and as much as 50\% of the sources have a star formation rate surface density above 0.1 M$_{\sun}$ yr$^{-1}$ kpc$^{-2}$ which has been shown to be a threshold for star formation rate winds \citep{Heckman:2001ve}. It thus seems as if a large fraction of AGN and perhaps virtually all have nuclear starbursts capable of driving winds. 

\acknowledgements
We thank the referee for their valuable comments and suggestions. We thank Alberto Bollato, Sylvain Veilleux, Len Cowie and Amy Barger for significant discussions and feedback that greatly improved this paper.

\bibliographystyle{apj}

\end{document}